\documentstyle[12pt]{article}

\newcommand{\be}{\begin{equation}}
\newcommand{\ee}{\end{equation}}

\newcommand{\prt}{\partial}

\newcommand{\bt}{\beta}
\newcommand{\vp}{\varphi}
\newcommand{\ep}{\varepsilon}
\newcommand{\al}{\alpha}
\newcommand{\ra}{\rightarrow}

\newcommand{\Gm}{\Gamma}

\begin{document}

\begin{center}

{\Large{\bf Optimization of self-similar factor approximants} \\ [5mm]

               V.I. Yukalov$^{a*}$ and S. Gluzman$^b$ } \\ [3mm]

{\it 
$^a$Bogolubov Laboratory of Theoretical Physics, \\
  Joint Institute for Nuclear Research, Dubna 141980, Russia \\ [3mm]

$^b$Generation 5 Mathematical Technologies Inc., Corporate Headquarters, \\
         515 Consumer Road, Toronto, Ontario M2J 4Z2, Canada }

\end{center}

\vskip 2cm

\begin{abstract}

The problem is analyzed of extrapolating power series, derived for 
an asymptotically small variable, to the region of finite values of 
this variable. The consideration is based on the self-similar 
approximation theory. A new method is suggested for defining the odd 
self-similar factor approximants by employing an optimization 
procedure. The method is illustrated by several examples having the 
mathematical structure typical of the problems in statistical and 
chemical physics. It is shown that the suggested method provides a 
good accuracy even when the number of terms in the perturbative 
power series is small. 
\end{abstract}

\vskip 2cm


\vskip 3cm

$^*$Corresponding author

\newpage

\section{Introduction}

The problem of extrapolating power series is constantly met 
in many branches of physics and applied mathematics, when 
one considers realistic systems with strong interactions. 
The theoretical description of such systems can be realized 
by invoking some numerical techniques. For instance, the 
investigation of model fluids can be done by molecular 
dynamics simulations, as is reviewed in Ref. [1]. However, 
it is always desirable to have analytical equations allowing 
for an explicit description of the matter properties. The 
standard way of obtaining analytical equations is by using 
some kind of perturbation theory. But the difficulty in 
deriving such analytical equations is that the realistic matter 
not often possesses small parameters, with respect to which one 
could develop perturbative expansions. Vice versa, the usual 
situation is the absence of such parameters. It is possible, 
nevertheless, to formally employ perturbation theory with 
respect to some system parameters, keeping in mind that the 
resulting series necessarily require an effective resummation. 
Among the  methods, defining an effective limit of asymptotic 
series, the most popular are the Pad\'{e} approximation [2] 
and the Borel summation [3]. But, as is well known, the method 
of Pad\'{e} approximants is not uniquely defined and is plagued 
by the occurrence of annoying divergences. While the Borel 
summation is applicable only for the cases enjoying a large 
number of perturbative terms and the knowledge of the 
high-order expansion coefficients, which is rarely available. 
Recently, a novel general approach has been developed for 
extrapolating perturbative series, called the self-similar 
approximation theory [4-13]. It has been persuasively 
demonstrated by numerous examples [4-13] that this approach 
is much simpler and more accurate than other extrapolation 
techniques. 

The term {\it self-similar} with respect to the considered 
approximation theory is used because of the basic idea behind 
this approach. This idea can be described as follows. Suppose 
that by employing perturbation theory, we derive a sequence 
$\{f_k\}$ of approximations $f_k$, with the index $k = 1,2...$ 
enumerating the approximation order. Considering the index $k$ 
as the variable similar to discrete time, we can treat the 
motion from the $k$-th approximation term to the higher-order 
terms, $k+1, k+2,...$ as an evolution of a dynamical system.
A dynamical system with discrete time is called {\it cascade}.
Then the sequence $\{f_k\}$ represents the points of a cascade,
which form the cascade trajectory. If we would know the equation 
of motion for this dynamical system, then, starting from the 
initial points of the trajectory with low-orders $k = 1,2,...$,
we would be able to find the higher-order terms, including the 
point representing an effective limit $f^*$ of the sequence
$\{f_k\}$ as $k \ra \infty$. This effective limit corresponds 
to the fixed point of the cascade trajectory. In the vicinity
of the fixed point, it is possible to write down the equation 
of motion as a functional relation between the terms 
$f_{k+p}$ and $f_k$, where $p \geq 0$. This relation has the 
form of the renormalization-group equation, which is also called
the {\it equation of functional self-similarity} or the 
{\it relation of functional self-similarity}. Our approximation 
theory is based on the use of this self-similarity relation, 
because of which the method is named the {\it self-similar 
approximation theory}. All mathematical details and foundations 
of this approach can be found in Refs. [4-12].  

A very convenient and powerful variant of the self-similar 
approximation theory is the method of self-similar factor 
approximants [14-18]. The direct use of this method requires 
the knowledge of even numbers of terms in the asymptotic series. 
However, it would be certainly desirable to be able to work with 
any number of terms, whether even or odd, in order to use in full 
all the information contained in the given series. A simple way 
of avoiding this problem was employed in Refs. [16-18], where 
it was suggested to set one of the parameters of the odd factor 
approximants to unity, which could be explained by scaling 
arguments.

In the present paper, we propose a more justified method for 
defining the odd factor approximants. This method is based on the 
optimization procedure, which follows the ideas of the optimized 
perturbation theory [19,20] (see also review articles [21,22]). We 
demonstrate by a number of examples that this new method provides a 
good accuracy of extrapolation for various problems typical of 
statistical and chemical physics. The method works well even when 
the number of the terms in the asymptotic series is rather small. 
Comparing the results of the calculations, employing the suggested 
optimization procedure, with the results of the method using the 
scaling arguments [16-18] makes it possible to justify the latter 
method.

\section{Formulation of problem}

Suppose we aim at finding a real function $f(x)$ of the real 
variable $x$. But the equations, defining this function, allow us 
to derive its form only for asymptotically small $x$, where the 
sought function 
\be
\label{1}
f(x) \simeq f_k(x) \qquad (x \ra 0)
\ee
is approximated by the series
\be
\label{2}
f_k(x) = f_0(x) \sum_{n=0}^k a_n x^n  \; .
\ee
Here $f_0(x)$ is a term defined so that $a_0 = 1$.

The extrapolation of this asymptotic series to finite values of 
the variable $x$ can be done by means of the self-similar factor 
approximants [14-18] prescribed by the expression
\be
\label{3}
f_k^*(x) = f_0(x) \prod_{i=1}^{N_k} (1 + A_i x)^{n_i}\; ,
\ee
with the number of factors 
\begin{eqnarray}
\label{4}
N_k = \left \{ \begin{array}{ll}
k/2 , & k = 2,4, \ldots \\
(k+1)/2, & k = 3,5, \ldots 
\end{array} \right. \; .
\end{eqnarray}
The parameters $A_i$ and powers $n_i$ are defined by the 
accuracy-through-order procedure that means the following.
We expand the factor approximant (3) in powers of the 
variable $x$ up to the $k$-th order. The resulting 
expansion is compared with the initial series (2) and the 
coefficients at like orders of $x$ are equated with each 
other. Or we may compare the expansions of $k$-th order 
obtained from $\ln f_k(x)$  and $\ln f^*_k(x)$. Both these 
ways are evidently equivalent, since the logarithm of a 
function and the function itself are in one-to-one 
correspondence with each other. The second way gives 
slightly more convenient equations. Following this way, 
with equating the like-order terms in the expansions for
$\ln f_k(x)$  and $\ln f^*_k(x)$, we come to the system of
equations
\be
\label{5}
\sum_{i=1}^{N_k} \; n_i A_i^n = B_n \qquad (n=1,2,\ldots,k)\; ,
\ee
in which
$$
B_n = \frac{(-1)^{n-1}}{(n-1)!} \;
\lim_{x\ra 0} \; \frac{d^n}{dx^n} \ln \left (
\sum_{m=0}^n a_m x^m \right ) \; .
$$    
All values of $B_n$ are expressed through the coefficients 
$a_1, a_2,..., a_n$ of expansion (2).

Explicitly, Eqs. (5), e.g., for even $k$, are
$$
n_1 A_1 + n_2 A_2 + \ldots + n_{k/2} A_{k/2} = B_1   \; ,
$$ 
$$
n_1 A_1^2 + n_2 A_2^2 + \ldots + n_{k/2} A_{k/2}^2 = B_2   \; ,
$$ 
$$
\dots \ldots \ldots 
$$
$$
n_1 A_1^k + n_2 A_2^k + \ldots + n_{k/2} A_{k/2}^k = B_k   \; .
$$ 
The solutions to these equations involve Vandermonde determinants.
When $k$ is even, there are $k$ unknowns, $k/2$ parameters $A_i$ 
and $k/2$ powers $n_i$. Then all these unknowns are uniquely 
defined. But when $k$ is odd, there are $k+1$ unknowns, that is, 
$(k+1)/2$ parameters $A_i$ and $(k+1)/2$ powers $n_i$. Then, to 
define uniquely all parameters, it is necessary to impose an 
additional condition defining, for instance, $A_1$. In Refs. 
[16-18], we used the condition $A_1 = 1$, following from scaling 
arguments. In the present paper, we suggest another, more general 
and justified condition, based on optimization procedure.

\section{Optimization method}

The idea of the method is to choose $A_1 \equiv A$ so that this 
choice would not influence much all other parameters $A_i$ and 
$n_i$. In other words, the factor approximant (3) should weakly 
depend on the chosen $A$. For small $x$, approximant (3) tends 
to
\be
\label{6}
f_k^*(x) \simeq f_0(x) \qquad (x \ra 0)  \; ,
\ee
which does not depend on $A$, hence
$$
\frac{\prt f_k^*(x)}{\prt A} \simeq 0 \qquad ( x \ra 0)\; .
$$   
For large $x$, approximant (3) behaves as
\be
\label{7}
f_k^*(x) \simeq f_0(x) B x^\bt \qquad (x\ra \infty) \; ,
\ee
where
\be
\label{8}
 B \equiv \prod_{i=1}^{N_k} A_i^{n_i} \; , \qquad
\bt \equiv \sum_{i=1}^{N_k} n_i \; ,  
\ee
and $N_k = (k+1)/2$. Differentiating form (7) over $A$ yields
$$
\frac{\prt }{\prt A} \; f_k^*(x) \simeq f_0(x) x^\bt \left (
\frac{\prt B}{\prt A} + B \ln x \; \frac{\prt \bt}{\prt A}
\right ) \;  .
$$
The solutions to Eqs. (5) are expressed through the unknown 
$A = A_1$, so that $A_i = A_i(A)$ and $n_i = n_i(A)$. From 
Eqs. (8) it follows that $B = B(A)$ and $\beta = \beta(A)$.
Inverting these equations gives $A = A(B)$ and $A = A_1(\beta)$.
Equating $A(B) = A_1(\beta)$ implies that $\beta = \beta (B)$, 
and, hence, $\beta = \beta(B(A))$. Therefore we have
\be
\label{9}
\frac{\prt }{\prt A} \; f_k^*(x) \simeq f_0(x) x^\bt 
\left ( 1 + B \ln x \; \frac{\prt \bt}{\prt A} \right ) 
\frac{\prt B}{\prt A} \; .
\ee
In order that the factor approximant (7) be weakly sensitive 
with respect to $A$, it sufficient to set
\be
\label{10}
\frac{\prt B}{\prt A} = 0 \;  .
\ee
This is the {\it optimization condition} defining $A$, which we 
shall employ.

\section{Illustration of method}

Many realistic problems are so complicated that only a few terms
of series (2) can be derived by means of perturbation theory. 
Therefore, illustrating the suggested optimization method, we 
shall concentrate our attention on the case of short series (2),
containing just a few terms. We shall find the optimized factor
approximant
\be
\label{11}
f_3^{opt}(x) = f_0(x) ( 1 + Ax)^{n_1} (1 + A_2 x)^{n_2} \; ,
\ee
in which $A$ is prescribed by the optimization condition (10).
And we shall compare form (11) with the nonoptimized variants of 
the factor approximants
\be
\label{12}
f_3^*(x) = f_0(x) ( 1 + x)^{n_1} (1 + A_2 x)^{n_2} \; ,
\ee
and 
\be
\label{13}
f_4^*(x) = f_0(x) ( 1 + A_1x)^{n_1} (1+ A_2 x)^{n_2} \; ,
\ee
in which all parameters are defined by the accuracy-through-order 
procedure, that is, are given by Eqs. (5).

\subsection{Logarithmic function}

To show that the method of factor approximants can approximate 
transcendental functions, let us start with a logarithmic 
function
\be
\label{14}
f(x) = \frac{1}{x} \; \ln(1 + x) \; .
\ee
Its expansion in series (2) yields
$$
a_n = \frac{(-1)^n}{n+1} \; , \qquad
f_0(x) = 1 \; .
$$
The first several coefficients explicitly are
$$
a_1 = - \;\frac{1}{2}, \qquad a_2 = \frac{1}{3}, \qquad
a_3 = - \;\frac{1}{4}, \qquad a_4 = \frac{1}{5}.
$$

For the optimized factor approximant (11), we find
$$
A = 0.65513, \qquad A_2 = 0.94080, \qquad n_1 = - 0.272, 
\qquad n_2 = -0.440.
$$
Its large-variable behavior is 
$$
f_3^{opt}(x) \simeq 1.405 x^{-0.712} \;  .
$$
This can be compared with the behavior of the nonoptimized 
approximant
$$
f_3^*(x) \simeq 1.260 x^{-0.667} \; .
$$
Note that the direct consideration of the asymptotic series (2)
for large $x$ has no sense, leading to either $- \infty$ or 
$+ \infty$, as $x \ra \infty$.

The factor approximants extrapolate the asymptotic series (2) to
the region of large variables. As an example, let us consider 
the accuracy of approximants (11), (12), and (13) for $x = 100$.
For $x = 100$, we have
$$
f_3^{opt}(100) = 0.052 \; , \qquad  f_3^*(100) = 0.058 \; , 
\qquad f_4^*(100) = 0.054 \; .
$$
Comparing this with the exact value $f(100) = 0.046$, we find the 
corresponding percentage errors
$$
\ep \left ( f_3^{opt} \right ) = 13\% \;, \qquad  
\ep \left ( f_3^* \right ) = 26\% \;, \qquad 
\ep \left ( f_4^* \right ) = 17\% \;  .
$$
As is seen, the optimized approximant is more accurate than the 
nonoptimized ones.

\subsection{Factorial function}

Factor approximants can well extrapolate even rather complicated 
transcendental functions, such as the factorial function
\be
\label{15}
f(x) = \frac{1}{\sqrt{2\pi} } \; e^{1/x} x^{1/2+1/x}
\Gm\left ( 1 + \frac{1}{x} \right ) \;  ,
\ee
where $\Gamma(x)$ is the gamma function. In expansion (2) for 
this function, one has $f_0(x) = 1$ and
$$
a_1 = \frac{1}{12}, \qquad  a_2 = \frac{1}{288}, \qquad
a_3 = - \;\frac{139}{51840}, \qquad a_4 = -\; \frac{571}{2488320}.
$$
The large-variable behavior of function (15) is
\be
\label{16}
f(x) \simeq 0.398942 x^{0.5} \qquad (x\ra\infty) \; .
\ee

For the optimized approximant (11), we have
$$
A = 0.31614, \qquad A_2 = 0.31630, \qquad 
n_1 = 474.945, \qquad n_2 = - 474.418.
$$
The large-variable form of approximant (16) is
$$
 f_3^{opt}(x) \simeq 0.419 x^{0.527} \qquad (x\ra\infty) \;  .
$$ 
This can be compared with the large-variable behavior of the 
nonoptimized approximant
$$
 f_3^*(x) \simeq 0.119 x^{0.917} \qquad (x\ra\infty) \;  .
$$
Again, the optimized factor approximant (11) essentially more 
accurately extrapolates the asymptotic series (2). For instance,
at $x = 100$, the error of the optimized approximant is about 
$15\%$, while the errors of the nonoptimized ones are several 
times larger.

\subsection{Debye-H\"{u}ckel function}

In the Debye-H\"{u}ckel theory of strong electrolytes [23], there 
arises the function
\be
\label{17}
D(x) = \frac{2}{x} \; - \; \frac{2}{x^2} \left ( 1 - e^{-x}
\right ) \; .
\ee
The related series (2) contain $f_0(x) = 1$ and the coefficients
$$
a_1 = -\; \frac{1}{3}, \qquad a_2 = \frac{1}{12}, \qquad 
a_3 = - \; \frac{1}{60}, \qquad a_4 = \frac{1}{360}.
$$
At large $x$, one has
\be
\label{18}
D(x) \simeq 2 x^{-1} \qquad (x \ra \infty) \; .
\ee

For the optimized approximant (11), we get
$$
A = 0.2965, \qquad A_2 = 0.2950, \qquad  n_1 = 98.602, 
\qquad n_2 = - 100.221.
$$
Approximants (11), (12), and (13) extrapolate reasonably well 
series (2) up to $x\sim 10$. Comparing the accuracy of the 
approximants at $x = 10$, we get the percentage errors
$$
\ep\left(D_3^{opt} \right ) = -13\% \; , \qquad
\ep\left(D_3^* \right ) = -18\% \; , \qquad 
\ep\left(D_4^* \right ) = -5 \% \; .
$$  
But for larger $x \gg 10$, the accuracy of all these
approximants becomes not sufficient.

\subsection{Statistical integrals}

In the problems of statistical physics [23], there occur the 
integrals of the following type
\be
\label{19}
I(x) = \int_0^\infty \; \frac{e^{-u} }{1+xu} \;du \; .
\ee
Expanding this in powers of $x$, one has series (2), with 
$f_0(x) = 1$ and the coefficients
$$
a_1 = - 1, \qquad a_2 = 2, \qquad a_3 = - 6, \qquad a_4 = 24.
$$
Such a series is strongly divergent. In the limit of large $x$,
the behavior of integral (19) is as $(\ln x)/x$, similar to that 
of the logarithmic function (14).

For the optimized approximant (11) we find
$$
A = 2.936042, \qquad A_2 = 65.541043, \qquad n_1 = - 0.340, 
\qquad n_2 = - 1.559 \times 10^{-5}.  
$$
To compare the accuracy of the extrapolation, let us consider 
the value of $x =10$. The corresponding percentage errors are
$$
\ep\left(I_3^{opt} \right ) = 56\% \; , \qquad
\ep\left(I_3^* \right ) = 1\% \; , \qquad 
\ep\left(I_4^* \right ) = 20 \% \;   .
$$ 
In this case, the most accurate extrapolation is achieved by 
means of the factor approximant $I_3^*(x)$.

A close type of the integral, met in some problems of 
statistical physics [23], reads as
\be
\label{20}
I(x) = (1 + 2x) \int_0^\infty \; 
\frac{e^{-u} }{1+x^2u^2} \;du \; .
\ee
Its expansion in series (2) gives $f_0(x) = 1$ and the 
coefficients
$$
a_1 = 2, \qquad a_2 = - 2, \qquad a_3 = - 4,
\qquad a_4 = 24.
$$
This series are also strongly divergent.

For the optimized approximant (11), we have
$$
A = 3.370958 - 3.406509\; i, \qquad A_2 = A^*, \qquad
n_1 = 0.119 + 0.175\; i, \qquad n_2 = n_1^*.
$$
Comparing different approximants at the point $x = 5$, we 
find their percentage errors
$$
\ep\left(I_3^{opt} \right ) = 12\% \; , \qquad
\ep\left(I_3^* \right ) = 5\% \; , \qquad 
\ep\left(I_4^* \right ) = 18 \% \;   .
$$
Here again, the most accurate extrapolation is provided by
the approximant $I_3^*(x)$, although all three approximants 
(11), (12), and (13) have errors of the same order of 
magnitude.

\subsection{Partition function}

The typical structure of partition functions in statistical 
physics [23] is well described by the integral
\be
\label{21}
Z(g) = \frac{1}{\sqrt{\pi}} \; \int_{-\infty}^\infty \;
\exp\left ( -\vp^2 - g\vp^4 \right ) \; d\vp \;  ,
\ee
corresponding to the zero-dimensional field model, where 
$g > 0$ plays the role of the coupling parameter. Expanding 
Eq. (21) in powers of $g$ yields series (2), with $g$ instead 
of $x$, with $f_0(g) = 1$, and the coefficients
$$
a_n = \frac{(-1)^n}{\sqrt{\pi}\; n!}\; \Gm\left ( 2n +
\frac{1}{2} \right ) \; .
$$
The first coefficients explicitly are
$$
a_1 = -\; \frac{3}{4}, \qquad a_2 = \frac{105}{32}, \qquad
a_3 = - \; \frac{3465}{128}, \qquad a_4 = \frac{675675}{2048}.
$$
The series is strongly divergent for any finite $g$. In the 
limit of large $g$, one has
\be
\label{22}
Z(g) \simeq 1.023 g^{-0.25} \qquad
(g\ra \infty) \; .
\ee
For the optimized approximant (11), we find
$$
A = 3.678882, \qquad A_2 = 16.099756, 
\qquad n_1 = - 0.132943, \qquad n_2 = - 0.016206.
$$
In the strong-coupling limit, we get
$$
 Z_3^{opt}(g) \simeq 0.804 g^{-0.149} \qquad
(g\ra \infty) \; .  
$$ 
In this limit, the approximants (12) and (13), respectively, are
$$
 Z_3^*(g) \simeq 0.917 g^{-0.346} \; , \qquad 
Z_4^*(g) \simeq 0.806 g^{-0.149} \qquad
(g\ra \infty) \;   .
$$
The accuracy of the approximants can be characterized by 
considering how well they approximate the amplitude and power
of the strong-coupling limit, as compared to the exact 
behavior (22). We find that the amplitude is approximated with 
the percentage errors  
$$
\ep\left(Z_3^{opt} \right ) = - 21\% \; , \qquad
\ep\left(Z_3^* \right ) = - 10\% \; , \qquad 
\ep\left(Z_4^* \right ) = - 21 \% \; ,
$$
while the power is predicted within the percentage errors
$$
 \ep\left(Z_3^{opt} \right ) = - 40\% \; , \qquad
\ep\left(Z_3^* \right ) = 38\% \; , \qquad 
\ep\left(Z_4^* \right ) = - 48 \% \;  .
$$
As is seen, the errors are of the same order of magnitude. 
Though these errors are not very small, we should not forget 
that the strong-coupling limit is predicted here on the basis
of only weak-coupling terms, where $g \ra 0$, with no assumption 
on the large $g \ra \infty$ behavior. Taking into account such 
a scarce information on just four weak-coupling terms, the 
prediction of the strong-coupling limit is not so bad.

\subsection{Anharmonic oscillator}

A great variety of problems in atomic and molecular physics, 
as well as in statistical physics and field theory, are modeled 
by the anharmonic oscillator, in which the anharmonicity 
corresponds to particle interactions. The anharmonic oscillator is
represented by the Hamiltonian
\be
\label{23}
H = -\; \frac{1}{2} \; \frac{d^2}{dx^2} + 
\frac{1}{2} \; x^2 + g x^4 \; ,
\ee
where $x \in (-\infty, \infty)$ and $g \in [0, \infty)$. This 
model has been extensively studied by perturbation theory with 
respect to the coupling parameter $g$. A detailed review article,
describing the results of perturbation theory, is Ref. [24]. 
However, perturbation theory for the energy levels yields strongly 
divergent series for any finite value of the coupling parameter 
$g$. Semiclassical approximation [24] can be used only for highly 
excited levels, but is not correct for low energy levels. In order 
to find effective limits of divergent series, the optimized 
perturbation theory [19,20], has been advanced, where the optimization 
is provided by introducing control functions governing the series 
convergence. The models of anharmonic crystals have been treated in 
this way [25-27]. Different variants of the optimized perturbation 
theory for anharmonic oscillators have been considered [25-33]. The 
use of the optimized perturbation theory requires the introduction 
of control functions into the starting zero-order approximation. 
These control functions are to be defined at each approximation order, 
which makes calculations at high orders rather complicated. Our aim 
in the present paper is to show that the self-similar factor 
approximants can also be applied to the anharmonic-oscillator problem. 
The advantage of the factor approximants is that they can be 
constructed on the basis of the pure pertubative series, without 
introducing control functions, which makes their use simpler.   
    
Calculating the ground state energy $e(g)$ by means of 
perturbation theory [11,24] yields series (2) in powers of $g$, with 
$$
f_0(g) = a_0 = \frac{1}{2}
$$
and with the coefficients
$$
a_1 = \frac{3}{4}, \qquad a_2 = - 2.625, \qquad a_3 = 20.8125, 
\qquad a_4 = - 241.2890625.
$$
The strong-coupling limit of the ground-state energy is
\be
\label{24}
e(g) \simeq 0.667986 g^{1/3} \qquad (g\ra\infty) \; .
\ee

For the optimized approximant (11), we obtain
$$
A = 3.79054, \qquad A_2 = 14.657819, \qquad 
n_1 = 0.224, \qquad n_2 = 0.044.
$$
In the strong-coupling limit, we have
$$
e_3^{opt}(g) \simeq 0.759 g^{0.269} \qquad (g\ra\infty) \; .
$$
Other approximants, in this limit, give
$$
e_3^*(g) \simeq 0.611 g^{0.590} \; ,\qquad 
e_4^*(g) \simeq 0.755 g^{0.231} \qquad (g\ra\infty) \; .   
$$
The amplitude of the strong-coupling limit is predicted by 
different approximants within the errors
$$
 \ep\left(e_3^{opt} \right ) = 13\% \; , \qquad
\ep\left(e_3^* \right ) = -8.5\% \; , \qquad 
\ep\left(e_4^* \right ) = 13 \% \; ,
$$
and the power of this limit is predicted within the errors
$$
 \ep\left(e_3^{opt} \right ) = -19\% \; , \qquad
\ep\left(e_3^* \right ) = 77\% \; , \qquad 
\ep\left(e_4^* \right ) = -31 \% \; .
$$
Again, we should not forget that these predictions for large 
$g \ra \infty$ are based on the very scarce information of 
only four terms of the weak-coupling expansion, when $g \ra 0$.
Going to the higher-order approximants essentially improves the 
accuracy. But let us recall that in the present paper we are 
interested in the possibility of extrapolating the weak-variable 
expansions to the large-variable range, when just a few first 
terms of such expansions are available.

\subsection{Critical indices}

In this section, we demonstrate the applicability of the method 
to calculating the critical indices related to the second-order 
phase transitions. We consider the $\varphi^4$ - field theory, 
for which the critical indices can be calculated by means of the 
Wilson $\varepsilon$-expansion, where $\varepsilon = 4 - d$, 
with $d$ being the space dimensionality (see, e.g., [34]). Let us 
be interested in the critical index $\nu = \nu(\varepsilon)$, 
which, under the $\varepsilon$-expansion, is a function of 
$\varepsilon$. One usually obtains the expansion for the function
\be
\label{25}
f(\ep) = \frac{1}{2\nu(\ep)} \; .
\ee
Treating $\varepsilon$ as an asymptotically small parameter, one 
gets [34] the expansion
\be
\label{26}
f(\ep) \simeq 1 + a_1 \ep + a_2 \ep^2 +
a_3 \ep^3 + a_4\ep^4 \; ,
\ee
in which $\varepsilon \ra 0$ and the coefficients are
$$
a_1 = - 0.1665, \qquad a_2 = - 0.05865, \qquad 
a_3 = 0.06225, \qquad a_4 = - 0.1535.
$$
However, the physical value of the index $\nu$ corresponds to 
the space dimensionality $d = 3$, hence, to the value $\ep = 1$. 
This means that it is necessary to extrapolate the asymptotic 
series (26) to the finite values of $\varepsilon$. 

Accomplishing the extrapolation using the method of self-similar
factor approximants, we extrapolate, first, the asymptotic 
series (26) getting $f_k^*(\varepsilon)$. Then we set 
$\varepsilon = 1$. This allows us to define the corresponding 
approximation for the index
\be
\label{27}
\nu_k^* = \frac{1}{2f_k^*(1)}  \; .  
\ee

Following the scheme of Sec. 3, for the optimized approximant (11),
we find
$$
A = 0.423, \qquad A_2 = 0.425, \qquad n_1 = - 306.868, 
\qquad n_2 = 305.275.
$$
In different approximations, we obtain the values
$$
\nu_3^{opt} = 0.614 \; , \qquad \nu_3^* = 0.617 \; ,
\qquad \nu_4^*=0.634 \; .
$$
These values are very close to those derived by the Borel 
summation, resulting in
$$
\nu = 0.628 \pm 0.001 \;\; {\rm [35]} \; , \qquad
\nu = 0.629 \pm 0.003 \;\; {\rm [36]} \;   ,
$$
and lattice numerical calculations 
$$
\nu = 0.631 \pm 0.002 \;\; {\rm [37]} \;   .
$$ 
Contrary to the latter numerical techniques, the method of 
self-similar factor approximants is much more simple, giving 
the critical index of the close value of magnitude.

\subsection{Polymer chain}

The expansion factor for a three-dimensional polymer chain 
with excluded-volume interactions is denoted as $\alpha (z)$,
where $z$ is a dimensionless coupling parameter [38,39]. 
Perturbation theory [38] results in an asymptotic series of 
form (2), with $f_0(z) = 1$ and the coefficients
$$
a_1 = \frac{4}{3}, \qquad a_2 = - 2.075385396, \qquad 
a_3 = 6. 296879676, \qquad a_4 = - 25.05725072.
$$
Numerical fitting [38,39] shows that the polymer expansion 
factor can be represented by the empirical dependence
\be
\label{28}
\al(z) = \left ( 1 + 7.524 z + 11.06 z^2 \right )^{0.1772}
\ee
and in the strong-coupling limit, it behaves as 
\be
\label{29}
\al(z) \simeq 1.531 z^{0.3544} + 0.184 z^{-0.5756} \; .
\ee

For the optimized approximant (11), we have
$$
A = 11.778451, \qquad A_2 = 4.118898, \qquad 
n_1 = 4.84 \times 10^{-3}, \qquad n_2 = 0.31.
$$ 
The strong-coupling limit for this approximant is
$$
\al_3^{opt}(z) \simeq 1.569 z^{0.315} \qquad (z\ra\infty) \; .
$$
Approximants (12) and (13) give the strong-coupling limits
$$
\al_3^*(z) \simeq 1.424 z^{0.440} \; , \qquad
\al_4^*(z) \simeq 1.560 z^{0.340} \qquad (z\ra\infty) \;   .
$$

The strong-coupling limit here is of special importance, since 
it defines the strong-coupling exponent
\be
\label{30}
\nu \equiv \frac{1}{2} + \frac{1}{4} \; \lim_{z\ra\infty}
\; \frac{\ln \al(z)}{\ln z} \; ,
\ee
which is a kind of the critical index for the polymer chain [40].
It shows how the polymer expansion factor behaves at large 
coupling parameters,
$$
\al(z) \; \propto \; z^{2(2\nu-1)} \;  .
$$ 

For our approximants (11), (12), and (13), we find the index
$$
\nu_3^{opt} =0.579 \; , \qquad \nu_3^*=0.610 \; ,
\qquad \nu_4^*=0.585 \;  .
$$
This should be compared with the numerical estimates
$$
 \nu=0.5886\;\; {\rm [38,39]} , \qquad 
 \nu=0.5877 \pm 0.0006\;\; {\rm [40]} \;  .
$$
As is seen, the self-similar factor approximants yield the values
that are very close to these numerical results.

\section{Conclusion}

In the development of the method of self-similar factor 
approximants, we have suggested a solution to the problem of 
constructing the approximants of odd orders. We propose to 
define one of the parameters of the factor approximant by the 
optimization condition of form (10). The advanced technique
is illustrated by a number of different problems with the 
mathematical structure typical of statistical physics and 
chemical physics. In these problems, it is necessary to
extrapolate the asymptotic series, derived for an 
asymptotically small variable, to the finite values of the 
latter and often one is interested in extrapolating the series
to infinitely large variables. Calculations demonstrate that 
the method of self-similar factor approximants gives rather 
accurate results even when the number of terms in the 
asymptotic series is small.

It is worth emphasizing that the method of self-similar factor 
approximants allows for accurate approximations not only for 
rational functions, as in the method of Pad\'e approximants, 
but also for transcendental functions, as are considered in 
Secs. 4.1, 4.2, and 4.3. 

Moreover, there exists a wide class of functions that can be 
reconstructed exactly. This class of functions is formed by the 
following structures. Let $P_n(x)$ be a polynomial in a real 
variable $x$ of degrees $n$ over the field of real numbers. 
And let $\alpha_i$ and $\beta_j$ be arbitrary complex numbers. 
Compose the real-valued products of powers of such polynomials as
$$
\prod_i \; P_{m_i}^{\al_i}(x) \; , \qquad
\prod_j \; P_{n_j}^{\bt_j}(x) \;  .
$$      
The real values for these products are guaranteed, provided the 
complex powers $\alpha_i$ and $\beta_j$ enter the corresponding 
products in complex conjugate pairs. Introduce the function 
$R_{MN}(x)$ given by the ratio
$$
R_{MN}(x) \equiv 
\frac{\prod_i P_{m_i}^{\al_i}(x)}{\prod_j P_{n_j}^{\bt_j}(x)} \; ,
$$
where
$$
M = \sum_i m_i \; , \qquad N = \sum_j n_j \; .
$$
Denote the class of functions with the above structure through
$$
{\cal R} \equiv \{ R_{MN}(x) : \;
M \geq 0 , \; N \geq 0 \} \; .
$$ 
Suppose that a function $f(x)$ is represented by an asymptotic 
series (2). Define the function
$$
\vp(x) \equiv \frac{f(x)}{f_0(x)} \; .   
$$
Then the following theorem holds.

{\it Theorem}. The function $f(x)$, with expansion (2), is 
reproduced by means of the self-similar factor approximants (3),
for any order $k \geq M + N$, exactly if and only if the 
function $\varphi(x)$ pertains to the class $\cal R$ of functions
of the form $R_{MN}(x)$.

The proof of this theorem is given in Ref. [14] and also 
discussed in Ref. [15]. Concrete examples of the exact 
reconstructions are illustrated in Refs. [14-16,18].

In conclusion, we may note that if the perturbative series do 
not have the form of an expansion in integer powers, as in Eq. 
(2), it is often easy to reduce them to that form by a change of 
variables. For instance, if we have an expansion in powers of 
$\sqrt{x}$, by the change $x = g^2$ we come to the expansion in 
integer powers for $g$.

\vskip 5mm
{\bf Acknowledgement}

\vskip 3mm
We are grateful to E.P. Yukalova for numerous useful 
discussions. Financial support from the Russian Foundation 
for Basic Research is appreciated.


\newpage

\begin{thebibliography}{99}

\bibitem{1}
 D.M. Heyes and A.C. Branka, Mol. Phys. {\bf 107}, 309 (2009).            

\bibitem{2}
 G.A. Baker and P. Graves-Moris, {\it Pad\'{e} Approximants} 
 (Cambridge University, Cambridge, 1996).   

\bibitem{3}
 H. Kleinert, {\it Path Integrals} (World Scientific, Singapore, 
 2006).
  
\bibitem{4}
 V.I. Yukalov, Phys. Rev. A {\bf 42}, 3324 (1990).

\bibitem{5}
 V.I. Yukalov, Physica A {\bf 167}, 833 (1990).

\bibitem{6}
 V.I. Yukalov, J. Math. Phys. {\bf 32}, 1235 (1991).

\bibitem{7}
 V.I. Yukalov, J. Math. Phys. {\bf 33}, 3994 (1992).

\bibitem{8}
 V.I. Yukalov and E.P. Yukalova, Physica A {\bf 198}, 573 (1993).

\bibitem{9}
 V.I. Yukalov and E.P. Yukalova, Int. J. Mod. Phys. B {\bf 7}, 
 2367 (1993).  

\bibitem{10}
 V.I.Yukalov and E.P. Yukalova, Physica A {\bf 206}, 553 (1994).

\bibitem{11}
 V.I. Yukalov and E.P. Yukalova, Laser Phys. {\bf 5}, 154 (1995).

\bibitem{12}
 V.I. Yukalov and E.P. Yukalova, Physica A {\bf 225}, 336 (1996).

\bibitem{13}
 V.I. Yukalov and S. Gluzman, Phys. Rev. Lett. {\bf 79}, 333 (1997). 

\bibitem{14}
 S. Gluzman, V.I. Yukalov, and D. Sornette, Phys. Rev. E {\bf 67},
 026109 (2003). 

\bibitem{15}
 V.I. Yukalov, S. Gluzman, and D. Sornette, Physica A {\bf 328}, 
 409 (2003).

\bibitem{16}
 V.I. Yukalov and E.P. Yukalova, Phys. Lett. A {\bf 368}, 341 (2007).

\bibitem{17}
 V.I. Yukalov and E.P. Yukalova, Eur. Phys. J. B {\bf 55}, 93 (2007).

\bibitem{18}
 E.P. Yukalova, V.I. Yukalov, and S. Gluzman, Ann. Phys. (N.Y.) 
 {\bf 323}, 3074 (2008).  

\bibitem{19}
 V.I. Yukalov, Moscow Univ. Phys. Bull. {\bf 31}, 10 (1976).

\bibitem{20}
 V.I. Yukalov, Theor. Math. Phys. {\bf 28}, 652 (1976). 

\bibitem{21}
 V.I. Yukalov and E.P. Yukalova, Ann. Phys. (N.Y.) {\bf 277}, 219 (1999).

\bibitem{22}
 V.I. Yukalov and E.P. Yukalova, Chaos Solit. Fract. {\bf 14}, 
 839 (2002).

\bibitem{23}
 L.D. Landau and E.M. Lifshitz, {\it Statistical Physics}
 (Butterworth-Heinemann, Oxford, 2000).

\bibitem{24}
 F.T. Hioe, D. McMillen, and E.W. Montroll, Phys. Rep. {\bf 43},
 307 (1978).

\bibitem{25}
 V.I. Yukalov, Physica A {\bf 89}, 363 (1977).

\bibitem{26}
 V.I. Yukalov, Ann. Physik {\bf 36}, 31 (1979).

\bibitem{27}
 V.I. Yukalov, Ann. Physik {\bf 37}, 171 (1980).

\bibitem{28}
 W.E. Caswell, Ann. Phys. (N.Y.) {\bf 123}, 153 (1979).

\bibitem{29}
 I. Halliday and P. Suranyi, Phys. Rev. D {\bf 21}, 1529 (1980).

\bibitem{30}
 I.K. Dmitrieva and G.I. Plindov, Phys. Lett. A {\bf 79}, 47 (1980).

\bibitem{31}
 J. Killingbeck, J. Phys. A {\bf 14}, 1005 (1981).

\bibitem{32}
 P.M. Stevenson, Phys. Rev. D {\bf 23}, 2916 (1981).

\bibitem{33}
 A. Okopi\'nska, Phys. Rev. D {\bf 35}, 1835 (1987).

\bibitem{34}
 H. Kleinert and V. Schulte-Frohlinde, {\it Critical Properties
    of $\varphi^4$ - Theories} (World Scientific, Singapore, 2006).

\bibitem{35}
 S.G. Gorishny, S.A. Larin, and F.V. Tkachev, Phys. Lett. A 
    {\bf 101}, 120 (1984).

\bibitem{36}
 R. Guida and J. Zinn-Justin, J. Phys. A {\bf 31}, 8103 (1998).

\bibitem{37}
 J. Zinn-Justin, {\it Quantum Field Theory and Critical 
    Phenomena} (Clarendon, Oxford, 1996).
    
\bibitem{38}
 M. Muthukumar and B.G. Nickel, J. Chem. Phys. {\bf 80}, 5839
    (1984).

\bibitem{39}
 M. Muthukumar and B.G. Nickel, J Chem. Phys. {\bf 86}, 460
    (1987).

\bibitem{40}
 B. Li, N. Madras, and A.D. Sokal, J. Stat. Phys. {\bf 80},
    661 (1995).
\end{thebibliography}
\end{document}